%% file: main.tex
\documentclass[a4paper,11pt]{article} 

\usepackage{authblk}
\usepackage{array}
\usepackage{url}
\usepackage{graphicx}
\usepackage{xcolor}
\usepackage{amssymb}
\usepackage{amsthm}
\usepackage{amsmath}
\usepackage{fullpage}
\usepackage{algorithm}
\usepackage{algpseudocode}
\usepackage{tabularx}
\usepackage{booktabs}
\usepackage{lipsum}
\usepackage{bm}
\usepackage{tikz}
\usepackage{bbold}

\definecolor{airforce}{rgb}{0.16,0.32,0.75}
\definecolor{cobalt}{rgb}{0.0,0.28,0.67}

\usetikzlibrary{arrows.meta, positioning}
\usepackage[colorlinks,urlcolor=cobalt,citecolor=cobalt,linkcolor=cobalt,pdftex, pdfauthor = Lukas Exl]{hyperref}

\usepackage{caption}
\usepackage{subcaption}
\usepackage[section]{placeins}

\usepackage{authblk}
\usepackage{mathrsfs}
\usepackage[mathcal]{euscript}
\usepackage{bm}
\usepackage{mathabx}

\usepackage{tabularx}
\usepackage{setspace} % for \onehalfspacing
\usepackage{wasysym}

\usepackage{ulem}

\DeclareMathOperator{\err}{err}

\newcommand{\pdof}[2]{\frac{\partial #1}{\partial #2}}
\renewcommand{\vec}[1]{\boldsymbol{#1}}

\begin{document}

\title{\Large{\textbf{{Physics-informed machine learning and stray field computation with application to micromagnetic energy minimization}}}}
\author[1,2]{Sebastian Schaffer}
\author[3,4,1]{Thomas Schrefl}
\author[4]{Harald Oezelt}
\author[3,4]{Alexander Kovacs}
\author[3,4]{Leoni Breth}
\author[1,2]{Norbert J. Mauser}
\author[1,5]{Dieter Suess} 
\author[1,2]{Lukas Exl \thanks{Corresponding author: \texttt{lukas.exl@univie.ac.at}}} 

\affil[1]{\small Research Platform MMM Mathematics - Magnetism - Materials, University of Vienna, Vienna, Austria}
\affil[2]{Wolfgang Pauli Institute, Vienna, Austria}
\affil[3]{Christian Doppler Laboratory for magnet design through physics informed machine learning, University for Continuing Education Krems, Wr. Neustadt, Austria}
\affil[4]{Department of Integrated Sensor Systems, University for Continuing Education Krems, Wr. Neustadt, Austria}
\affil[5]{Faculty of Physics, University of Vienna, Vienna, Austria} 

\maketitle
\date

\noindent\textbf{Abstract.}
We study the full 3d static micromagnetic equations via a physics-informed neural network (PINN) ansatz for the continuous magnetization configuration. PINNs are inherently mesh-free and unsupervised learning models. In our approach we can learn to minimize the total Gibbs free energy with additional conditional parameters, such as the exchange length, by a single low-parametric neural network model. In contrast, traditional numerical methods would require the computation and storage of a large number of solutions to interpolate the continuous spectrum of quasi-optimal magnetization configurations. We also consider the important and computationally expensive stray field problem via PINNs, where we use a basically linear learning ansatz, called extreme learning machines (ELM) within a splitting method for the scalar potential. 
This reduces the stray field training to a linear least squares problem with precomputable solution operator. 
We validate the stray field method by means of numerical example comparisons from literature and illustrate the full micromagnetic approach via the NIST $\mu$MAG standard problem $\#3$. 

\noindent\textbf{Keywords.}
extreme learning machines, stray field computation, NIST standard problem $\#3$, conditional physics-informed neural networks, static micromagnetism, JAX\\

% 68T05 Computer science:  Artificial intelligence: Learning and adaptive systems 
% 62P35 Statistics: Applications: application to physics 
% 65Z05 Numerical analysis: application to physics

\newpage
%% \linenumbers

%% main text %%%%%%%%%%%%%%%%

\section{Introduction}

Micromagnetism is a field of study that focuses on the behavior of small magnetic systems, such as those found in permanent magnets \cite{fischbacher2018micromagnetics}, data storage devices \cite{bashir2012head} and sensors \cite{suess2018topologically, huber20173d}. One important problem in this field is energy minimization, that is finding the quasi-optimal configuration of a magnetic system. The total Gibbs free energy of a magnetic body $\Omega \subset \mathbb{R}^3$ is composed of four fundamental energy terms \cite{exl2020micromagnetism}, the stray field energy $E_s$, the Zeeman energy $E_{zee}$, the anisotropy energy $E_a$ and the exchange energy $E_{ex}$, i.e.
\begin{equation}\label{eq:totenergy}
    E = \int_\Omega \underbrace{-\frac{\mu_0}{2}M_s\vec m \cdot \vec H_s}_{E_s} \,\underbrace{-\mu_0 M_s\vec m \cdot \vec H}_{E_{zee}} + \underbrace{K_u (1 - (\vec m \cdot \vec a)^2)}_{E_a} + \underbrace{A_{ex}\|\nabla\vec m\|^2_F}_{E_{ex}} dx,
\end{equation}
where $\vec H_s$ is the stray field, $\vec H$ the external field and $\vec a$ the unit vector along the easy axis of the magnetic body. Further, $K_u$ is the uniaxial anisotropy constant, the exchange stiffness constant is denoted with $A_{ex}$, $\mu_0$ is the permeability of the vacuum and $M_s$ is the saturation magnetization.

One can for instance apply \textit{LaBonte's method} to minimize the energy and perform a curvilinear steepest descent optimization with the projected gradient of the energy to find a local minimum \cite{exl2014labonte}, or variants of nonlinear conjugate gradient methods \cite{fischbacher2017nonlinear,exl2019preconditioned}. This involves a finite difference or finite element discretization of the domain which often leads to a very small mesh size in order to accurately represent the geometry, stray field and other energy terms. 
In recent years, approaches were developed involving low-parametric models for the magnetization $\vec m(\vec x;\, \vec \omega)=\mathcal{N}(\vec x;\, \vec \omega)$. These models only depend on a small amount of parameters $\vec \omega$ where computations are performed with respect to $\vec \omega$ instead of the discretized spatial coordinates. Such models involve eigenmode developments obtained from the spectral decomposition of the effective field operator \cite{d2009spectral,perna2022computational} and multilinear (low-rank) tensor methods \cite{exl2014tensor}. The recently introduced data-driven neural network ansatz via physics-informed neural networks (PINN) \cite{raissi2019physics} in micromagnetism \cite{kovacs2022conditional,kovacs2022magnetostatics} can be seen as a low-parametric nonlinear model obtained form unsupervised learning from the underlying physics. Specifically, PINNs come with certain modelling and computational advantages such as smooth interpolation of the solutions and their derivatives \cite{hornik1989multilayer}, ability to model the solutions of high-dimensional parametric equations without the usual curse of dimensions, the usage of automatic differentiation, as well as their mesh-free and unsupervised nature.

In this work we use a new approach for the usually very expensive stray field computation with a physics-informed hard constrained extreme learning machine (ELM) \cite{huang2006extreme} ansatz. A splitting framework is used to avoid evaluation of the stray field outside the magnetic domain. We apply this method for full 3d micromagnetic energy minimization via a PINN ansatz for the magnetization configuration involving a conditional parameter. This is illustrated via the NIST $\mu$MAG Standard Problem $\#3$  \cite{mumag3}.

\section{Methods}\label{sec:methods}

\subsection{Physics-informed Neural Networks}
A feedforward neural network is a type of artificial neural network that consists of layers of interconnected nodes, where data flows through the network from the input layer to the output layer without loops. Each node represents a unit that performs a simple computation on the data, and the connections between nodes represent the weights that adjust the influence of the data on the computation. In a feedforward neural network, the output of each node is determined by the weighted sum of its inputs, passed through an activation function that determines the output of the node.
It can be described by a series of affine transformations and nonlinear activation functions, such as a sigmoid or exponential linear unit (elu), in order to generate a final output. This can be expressed as a mapping from an input space $\mathcal{X}$ to an output space $\mathcal{Y}$, e.g. $\mathcal{N}(\,.\,;\vec \omega) : \mathcal{X} \rightarrow \mathcal{Y}$. In more detail, a network with $T$ layers, which maps the input $\vec x$ to the output $\vec y$, can be written in a recursive way as 
\begin{equation}
\begin{split}
    \vec z^{(0)} &= \sigma^{(0)}(\vec x) \\
    \vec z^{(t)} &= \sigma^{(t)}(\vec W^{(t)}\vec z^{(t-1)} + \vec b^{(t)}) \quad \text{for}\; t=1, \dots, T,\\
    \vec z^{(T)} &= \vec y,
\end{split}
\end{equation}
where $\vec W^{(t)}$ and $\vec b^{(t)}$ are the weight matrix and bias vector of the $t$\textsuperscript{th} layer and   $\sigma^{(t)}$ is the respective activation function. The output of the $t$\textsuperscript{th} hidden layer is denoted with $\vec z^{(t)}$.The trainable parameters are summarized with $\vec \omega = \{(\vec W^{(t)}, \vec b^{(t)} )\,|\, t=1, \dots, T\}$.
With traditional supervised learning, the parameters of the network are adjusted during training in order to minimize an objective function. In a regression framework, this is often the mean squared error (MSE).
Given a training set with $n$ samples  $\mathcal S = \{(\vec x_i, \vec y_i)\,|\, i=1, \dots, n\} \subseteq \mathcal{X} \times \mathcal{Y}$, the neural network parameters $\vec \omega$ should minimize 
\begin{equation}
    \mathrm{MSE}(\vec \omega) = \frac{1}{|\mathcal S|} \sum_{(\vec x, \vec y) \in \mathcal S} \|\mathcal{N}(\vec x; \vec \omega) -  \vec y\|^2.
\end{equation}
In contrast, PINNs are trained in an unsupervised fashion. Here the training set only consists of the feature vectors $\mathcal{S} = \{\vec x_i\,|\, i=1, \dots, n\} \subseteq \mathcal{X}$ and the optimal trainable parameters are obtained by minimizing a function $\mathcal{L}(\vec w)$ which describes the underlying physics, e.g. the total energy $E = \int_\Omega \mathcal{E}\big(\vec m (\vec x) \big)\, dx$ with possible additional constraints in a penalty term, i.e.
\begin{align}\label{eq:pinn}
\mathcal{L}(\vec \omega) = \frac{1}{|\mathcal{S}|} \sum_{\vec x \in \mathcal{S}} \mathcal{E}\big(\mathcal{N}(\vec{x};\vec{\omega})\big) + \textrm{Penalty},
\end{align}
where $\mathcal E$ is the energy density.
 %In our case, we 
Training of neural networks is mostly done with first order methods such as stochastic gradient descent and automatic gradient computation with back-propagation. The loss function often involves partial derivatives with respect to the input. If the input is low-dimensional a forward mode automatic differentiation can be used efficiently. Hence, the gradient $\nabla_{\vec \omega}\mathcal{L}$ is computed with forward automatic differentiation followed by reverse mode, i.e. \textit{reverse-over-forward}.

\subsection{Physics-informed stray field computation}\label{sec:stay-field}
Stray field computation is a critical aspect of micromagnetism, as it involves determining the field produced by the magnetization of a magnetic system that extends beyond its boundaries. Accurate stray field computation is essential for understanding the behavior of magnetic systems,  enabling the design of efficient and effective devices and permanent magnets. The stray field problem is a whole space Poisson problem with the scalar potential $\phi$ decaying at infinity. The stray field is given by $\vec H_s = -\nabla \phi$, where the scalar potential fulfills \cite{carstensen1995adaptive,exl2018magnetostatic}

\begin{equation}\label{eq:stray-field}
    \begin{aligned}
    -\Delta \phi &= -\nabla\cdot \vec m  &\text{in} \; \Omega \subset \mathbb{R}^3, \\
    -\Delta \phi &= 0 &\text{in} \; \overline\Omega^c, \\
    [\phi] &= 0 &\text{on} \; \partial\Omega, \\
    \left[\pdof{\phi}{\vec n}\right] &= - \vec m \cdot \vec n &\text{on} \; \partial\Omega, \\
    \phi(\vec x) &= \mathcal{O}\left(\frac{1}{\|\vec x\|}\right) &\text{as} \; \|\vec x\|\rightarrow \infty.
    \end{aligned}
\end{equation}
The jump conditions $[\,\cdot\,]$ describe the behavior of the potential at the boundary of the magnet. 
The solution to \eqref{eq:stray-field} is given by \cite{abert2013numerical}
\begin{equation}
    \phi(\vec x) = -\frac{1}{4\pi}\left(\int_\Omega \frac{\nabla\cdot \vec m(\vec y)}{\|\vec x - \vec y\|}d\vec y
    - \int_{\partial\Omega} \frac{\vec m(\vec y)\cdot \vec n(\vec y)}{\|\vec x - \vec y\|}d\sigma(\vec y)\right).
\end{equation}
Traditional numerical methods, like the demagnetisation tensor method \cite{abert2013numerical,miltat2007numerical}, use such integral representations to compute the solution directly on tensorial computational grids. Finite element methods use the weak form of the differential equation \eqref{eq:stray-field} utilizing splitting of the potential in interior and exterior parts \cite{garcia2006adaptive,exl2014non,schrefl2007numerical}. In the following we will use a splitting ansatz but the strong form of \eqref{eq:stray-field}.

In the ansatz of \textit{Garcia-Cervera and Roma} \cite{garcia2006adaptive}, the potential is split into $\phi = \phi_1 + \phi_2$, where 
\[
\phi_1 = \bigg\{
\begin{array}{ll}
\phi_1^{int} & \text{in}\; \Omega\\ 
\phi_1^{ext} & \text{in}\; \overline\Omega^c\\ 
\end{array},\; \phi_2 = \bigg\{
\begin{array}{ll}
\phi_2^{int} & \text{in}\; \Omega\\ 
\phi_2^{ext} & \text{in}\; \overline\Omega^c\\ 
\end{array},
\]
where $\phi_1^{int},\; \phi_2^{int}\in  H^1(\Omega)$ and $\phi_1^{ext},\;\phi_2^{ext} \in H^1_{loc}(\overline\Omega^c)$. Inside the domain $\phi_1^{int}$ should satisfy

\begin{equation}
    \begin{aligned}
        -\Delta \phi_1^{int} &= - \nabla \vec m &\text{in}\; \Omega, \\ 
                \phi_1^{int} &= 0 &\text{on}\; \partial \Omega
    \end{aligned}
\end{equation}
and $\phi_1^{ext} = 0$ in $\overline\Omega^c$. Therefore, $[\phi_1] = 0$ and $[\pdof{\phi_1}{\vec n}] = - \pdof{\phi_1^{int}}{\vec n}$. In order for $\phi$ to fulfill \eqref{eq:stray-field}, $\phi_2$ has to satisfy

\begin{equation}\label{eq:phi2-interface}
    \begin{aligned}
        -\Delta \phi_2^{int} &= 0 &\text{in}\; \Omega, \\ 
        -\Delta \phi_2^{ext} &= 0 &\text{in}\; \overline\Omega^c, \\ 
        [\phi_2] &= 0 &\text{on}\; \partial \Omega, \\
        \left[\pdof{\phi_2}{\vec n}\right] &= -\vec m \cdot \vec n + \pdof{\phi_1^{int}}{\vec n} &\text{on}\; \partial \Omega, \\
        \phi_2^{ext}(\vec x) &= \mathcal{O}\left(\frac{1}{\|\vec x\|}\right) &\text{as} \; \|\vec x\|\rightarrow \infty.
    \end{aligned}
\end{equation}
The solution $\phi_2$ is given by the \textit{single layer potential},
\begin{equation}\label{eq:single-layer-pot}
    \phi_2(\vec x) = \frac{1}{4\pi}\int_{\partial\Omega} \frac{(\vec m\cdot \vec n - \pdof{\phi_1}{\vec n})(\vec y)}{\|\vec x - \vec y \|}\, ds(\vec y).
\end{equation}

One can use \eqref{eq:single-layer-pot} to compute $\phi_2$ in whole space, but in micromagnetism it is only required to find a solution inside the magnet(s) $\Omega$. The formulation in \eqref{eq:phi2-interface} allows to neglect $\phi_2^{ext}$ and only to solve the Laplace problem on the inner domain with the Dirichlet boundary conditions given by \eqref{eq:single-layer-pot}. The stray field problem can therefore be written without the exterior domain, as 
\begin{equation}\label{eq:garcia}
    \begin{aligned}
        -\Delta \phi_1 &= - \nabla \cdot \vec m &\text{in}\; \Omega, \\ 
                \phi_1 &= 0 &\text{on}\; \partial \Omega, \\
        -\Delta \phi_2 &= 0 &\text{in}\; \Omega, \\
              \phi_2(\vec x) &= \frac{1}{4\pi}\int_{\partial\Omega} \frac{(\vec m\cdot \vec n - \pdof{\phi_1}{\vec n})(\vec y)}{\|\vec x - \vec y \|}\, ds(\vec y) &\text{on}\; \partial \Omega \\[0.1cm]
              \phi^{int} &= \phi_1 + \phi_2 \\
              \vec H_s &= -\nabla \phi^{int}.
    \end{aligned}
\end{equation}

This approach has the advantage that it only requires the computation of a Dirichlet boundary problem with a unique solution. On the other hand, this formulation needs the evaluation of the normal derivative at the boundary. This can lead to inaccuracies in traditional mesh-based methods, but since neural networks are universal approximators it is possible to compute these derivatives with automatic differentiation up to machine precision.

We solve \eqref{eq:garcia} in a sequential way. First, a PINN is used for the homogeneous Dirichlet problem for $\phi_1$, followed by a separate PINN for $\phi_2$ with boundary conditions computed by the single layer potential \eqref{eq:single-layer-pot}. Instead of typical PINNs, as described above in \eqref{eq:pinn}, where the penalty accounts for the Dirichlet boundary conditions, we use a hard constrained ELM ansatz \cite{huang2006extreme}. It has been shown that these networks also possess universal approximation capabilities \cite{huang2006universal}. The hard constraints will always satisfy the boundary conditions leading to a more accurate description. This comes with several advantages: The objective is given by a linear least squares problem with exact solution and the solution operator can be precomputed. We can therefore avoid the usage of stochastic iterative optimization routines, which makes the computation very fast. This is important in the course of energy minimization since a huge amount of stray field evaluations is required.

A hard constraint formulation will always satisfy the given boundary or initial conditions. One such method is the theory of functional connections \cite{mortari2017theory}. This framework can be used to satisfy linear constraints on simple domains, but is difficult to apply to more complex geometries. If $\hat\phi$ is an approximate solution to the scalar potential, an alternative to this approach is the usage of the following ansatz
\begin{equation}\label{eq:hard_cons_formulation}
     \hat\phi(\vec x; \vec \omega) = \ell(\vec x)  h(\vec x; \vec \omega_1) +  g(\vec x; \vec \omega_2),
\end{equation}
with $\vec \omega = \{\vec \omega_1, \vec \omega_2\}$ being the parameters of two different models $h$ and $g$, where $h$ describes the physics of the interior domain and $g$ satisfies the boundary conditions. Further, the indicator function $\ell$ fulfills
\begin{equation}
    \ell(\vec x) = \bigg \{\begin{array}{ll}
0 & \text{if}\; \vec x \in \partial\Omega\\
> 0 & \text{if}\; \vec x \in \Omega.
\end{array}
\end{equation}
With this formulation, the Dirichlet boundary constraints are always satisfied if $g$ does. The solution to a Poisson problem can then be computed in three steps. First, find an indicator function $\ell$, second, compute the parameters $\vec \omega_2$ for the boundary model $g$ by minimizing the MSE and finally compute $\vec \omega_1$ by minimizing the final objective.

In order to construct the indicator function $\ell$, one can use the method described in \cite{sheng2021pfnn}. For that purpose, the boundary $\partial \Omega$ of the domain is partitioned into $K\geq 6$ segments $\{\partial \Omega_1, \dots, \partial \Omega_K\}$. For each segment $\partial \Omega_k$, select a non-adjacent segment $\partial \Omega_{k_o}$ to construct a spline function 
\begin{equation}
\left\{\begin{array}{ll}
l_k(\vec x) = 0 & \text{if}\; \vec x \in \partial\Omega_k,\\
l_k(\vec x) = 1 & \text{if}\; \vec x \in \partial\Omega_{k_o},\\
0 < l_k(\vec x) < 1& \mathrm{otherwise}.\\
\end{array}\right.
\end{equation}
The indicator function is then given by
\begin{equation}\label{eq:indicator_fun}
    \ell(\vec x) = \prod_{k=1}^K 1 - (1 - l_k(\vec x))^\mu,
\end{equation}
where the hyper-parameter $\mu \geq 1$ can adjust the shape of $\ell$. In \cite{sheng2021pfnn}, it is suggested to set $\mu = K$. 
For $l_k$ we use an inverse multiquadric radial basis interpolation \cite{wright2003radial, sheng2021pfnn}. For $M_k$ sample points $\{\vec x_1, \dots, \vec x_{M_k}\} \subset \partial \Omega_k \cup \partial \Omega_{k_o}$, $l_k $ is given by
\begin{equation}
    l_k(\vec x) = \sum_{i=1}^{M_k} a_i \frac{1}{\sqrt{(e^2 + \|\vec x - \vec x_i\|^2)}} + \vec b^T \vec x + c.
\end{equation}
The parameter $e$ is used to adjust the shape of the basis function. In \cite{sheng2021pfnn} it is recommended to set $e = 1.25 R / \sqrt{M_k}$ to maintain scale invariance, where $R$ is the radius of the minimal circle which encloses all sample points. The coefficients can be determined by solving the following linear system
\begin{equation}
\begin{aligned}
    \sum_{i=1}^{M_k} a_i \vec x_i &= \vec 0,\\
    \sum_{i=1}^{M_k} a_i  &= 0, \\
    l_k(\vec x_i) &=  \left\{\begin{array}{ll}
        0 & \text{if}\; \vec x_i \in \partial\Omega_k\\
        1 & \text{if}\; \vec x_i \in \partial\Omega_{k_o}\\
        \end{array}\right.,\,\, i = 1, \dots, M_k,
\end{aligned}    
\end{equation}
which has a unique solution if all sample points are distinct. This construction of the indicator function is applicable for complex geometries.

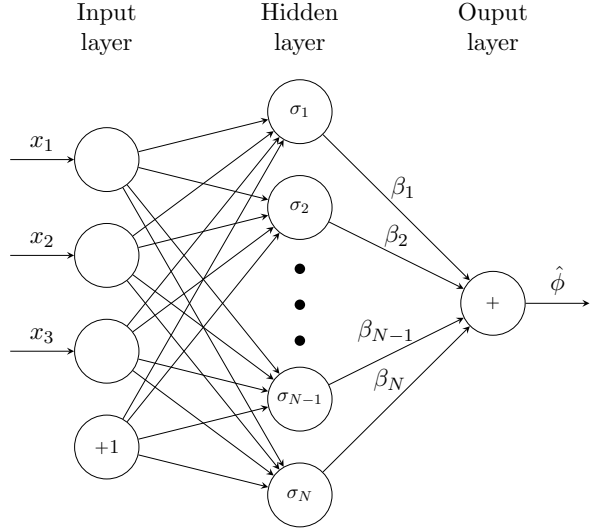
\begin{figure}
\center
\resizebox{0.5\textwidth}{!}{%
	\input{elm.tex}
}
\caption{\label{fig:elm} An extreme learning machine with $3$-dimensional input $\vec x$, $N$ hidden nodes and scalar output $\hat \phi$ where only the output parameter $\vec \beta$ is trainable.}
\end{figure}

For $h$ and $g$ we use ELMs. An ELM $\hat \phi$ can essentially be written as a linear combination of nonlinear activation functions, in our case
\begin{equation}\label{eq:elm}
    \hat\phi(\vec x; \vec\beta) = \sum_{i=1}^N \beta_i \sigma(\vec{w}_i^T\vec{x}+b_i) = \sum_{i=1}^N \beta_i \sigma_i(\vec{x}) \equiv \vec{\beta}^T \vec{\sigma}(\vec{x}).
\end{equation}
The only trainable parameter is $\vec\beta$ and the input weights are drawn from a random distribution, for the architecture see Fig.~\ref{fig:elm}.

Any linear operator applied on an ELM only acts on the activation functions and a linear combination of ELMs is again an ELM. In case of equation \eqref{eq:hard_cons_formulation}, where $h(\,\cdot\,;\vec \beta)$ and $g(\,\cdot\,;\vec \gamma)$ are ELMs, this takes the form
\begin{equation}\label{eq:elm_hard_constraints}
    \hat\phi(\vec x;\vec \beta,\vec\gamma) = \vec\beta^T\ell(\vec x)\vec\sigma_h(\vec x) + \vec\gamma^T\vec\sigma_g(\vec x),
\end{equation}
where $\ell(\vec x)\vec \sigma_h(\vec x)$ and $\vec \sigma_g(\vec x)$ are the respective hidden nodes. The application of the Laplace operator would result in 
\begin{equation}
    \Delta\hat\phi(\vec x;\vec \beta,\vec\gamma) = \vec\beta^T\Delta\left(\ell(\vec x)\vec\sigma_h(\vec x)\right) + \vec\gamma^T\Delta\vec\sigma_g(\vec x).
\end{equation}

For the Poisson problem $-\Delta \phi = f$ and boundary data $g^\ast(\vec x)$ with the penalty-free ELM ansatz $\hat \phi$, the training objective is given by
\begin{align}\label{eq:poisson_elm}
\int_\Omega |f(\vec x) + \Delta \hat \phi(\vec x; \vec\beta, \vec\gamma)|^2 \, dx \rightarrow \mathrm{min}_{{\vec\beta}}.
\end{align}
This is again computed by Monte Carlo integration and automatic differentiation. The boundary parameter $\vec \gamma$ can be precomputed in a supervised fashion from given exact values in order to minimize the mean squared error
\begin{equation}\label{eq:poisson_elm_bnd}
    \int_{\partial \Omega} |g(\vec x; \vec \gamma) - g^*(\vec x)|^2\, d s(\vec x) \rightarrow \mathrm{min}_{{\vec\gamma}}.
\end{equation}
Both equations \eqref{eq:poisson_elm} and \eqref{eq:poisson_elm_bnd} can be written as a linear least squares problem and the solution operator can be precomputed.

\subsection{Micromagnetic Energy Minimization with PINNs}

We intend to minimize the total Gibb's free energy \eqref{eq:totenergy} with a neural network ansatz for the magnetization. The stray field can be computed as described in Sec.~\ref{sec:stay-field} with precomputed solution operator for $\phi_1$ and $\phi_2$. In addition, we allow to include conditional parameters into the models in order to describe the whole family of solutions \cite{kovacs2022conditional}. Therefore, the sample set will not only include the spatial sample points, but also the conditional parameters, 
$\mathcal{S}=\{\vec{x}_1,\hdots,\vec{x}_{n_\Omega}\} \times \{\vec\lambda_1,\hdots,\vec\lambda_{n_\lambda}\} \subseteq \Omega \times \Lambda$, where $\Lambda \subset \mathbb{R}^c$ for $c$ conditional parameters. The total energy with penalty for the micromagnetic unit norm constraint with conditional neural network ansatz $\vec{m}(\vec{s};\vec{\omega}) = \mathcal{N}\big(\vec{s};\vec{\omega}\big),\,\, \vec{s}\in \mathcal{S}$ can be approximated with Monte Carlo integration 
\begin{align}\label{eq:objective_energy_min}
\frac{1}{|\mathcal{S}|} \sum_{\vec{s} \in \mathcal{S}} \mathcal{E}\big(\vec{m}(\vec{s},\vec{\omega})\big) + \alpha (\|\vec{m}(\vec{s},\vec{\omega})\|_2-1)^2\rightarrow \mathrm{min}_{\vec{\omega}}.
\end{align}
This formulation will compute the average total energy over the whole conditional domain. Here $\alpha > 0$ is the penalty parameter which is increased after several training epochs to enforce convergence to a constrained local optimum \cite{nocedal1999numerical}. Optimization of neural networks is usually performed with some variant of stochastic gradient descent (SGD). Since such an algorithmic framework often converges to the global minimum energy it is crucial to select a learning rate which is small enough in order to converge to a local optimum, which is often the desired goal for micromagnetic energy minimization. Further, stray field computation needs to be performed for every conditional parameter. The batch size of the conditional parameter should therefore be rather small to avoid a big computational load. 

Note that in this framework the stray field $\vec H_s(\,\cdot\,;\vec\omega) = - \nabla\phi^{int}(\,\cdot\,;\vec m)$, for some conditional parameter $\vec\lambda \in \Lambda$, is dependent on the model parameters $\vec \omega$ of the magnetization. Automatic differentiation can therefore also compute the gradient $\nabla_{\vec\omega} \vec H_s$ of the stray field. 
%This is not an issue and could be done in a similar fashion as in \cite{smith2010demagnetizing,bjork2021magtense}, but is an unnecessary step if the temperature is far from Curie temperature 
Hence we could model a nonlinear dependence on the magnetization in a similar fashion as in \cite{smith2010demagnetizing,bjork2021magtense}. Since the stray field is linear in $\vec m$, i.e. $\vec H_s([\vec x,\vec \lambda];\vec\omega) = \vec D(\vec x)\vec m([\vec x,\vec\lambda];\vec\omega)$, where $\vec D$ is the demagnetization tensor, the derivative, with respect to some parameter $\omega_i$, of the magnetostatic energy is then given by
\begin{equation}
\begin{aligned}
    \frac{\partial E_s}{\partial \omega_i} 
    &= -\frac{1}{2} \mu_0 M_s\frac{\partial}{\partial \omega_i} \int_\Omega \vec m \cdot \vec D \vec m\, dx \\
    &= -\mu_0 M_s\int_\Omega \vec H_s \frac{\partial \vec m}{\partial \omega_i} \, dx.
\end{aligned}
\end{equation}
Hence, gradient computation of the stray field with respect to the model parameter can be avoided.

\section{Results}
\subsection{Stray Field Computation}
In most cases the stray field cannot be expressed with analytical solutions.
However, a sphere with radius $r=1$ with magnetization $m=[0,0,1]$ produces the scalar potential $\phi^{int}(\vec x) = x_3 / 3$.  This case serves as a first test to validate the accuracy of our physics-informed stray field computation. The weights are drawn from a hypercube $[-1,1]^4$ using a Sobol sequence 
 \cite{sobol1976uniformly}. The same weights are used for all three ELM models and only the output parameter differs. We use 1024 samples drawn uniformly from the domain and 512 samples drawn uniformly from the boundary using a Sobol sequence as well. All activation functions are $\tanh$. For the indicator function we use $\ell(\vec x) = (r - \|\vec x\|_2 ^ 2 / r)$.
 \begin{figure}[ht]
     \centering
     \includegraphics[width=\textwidth]{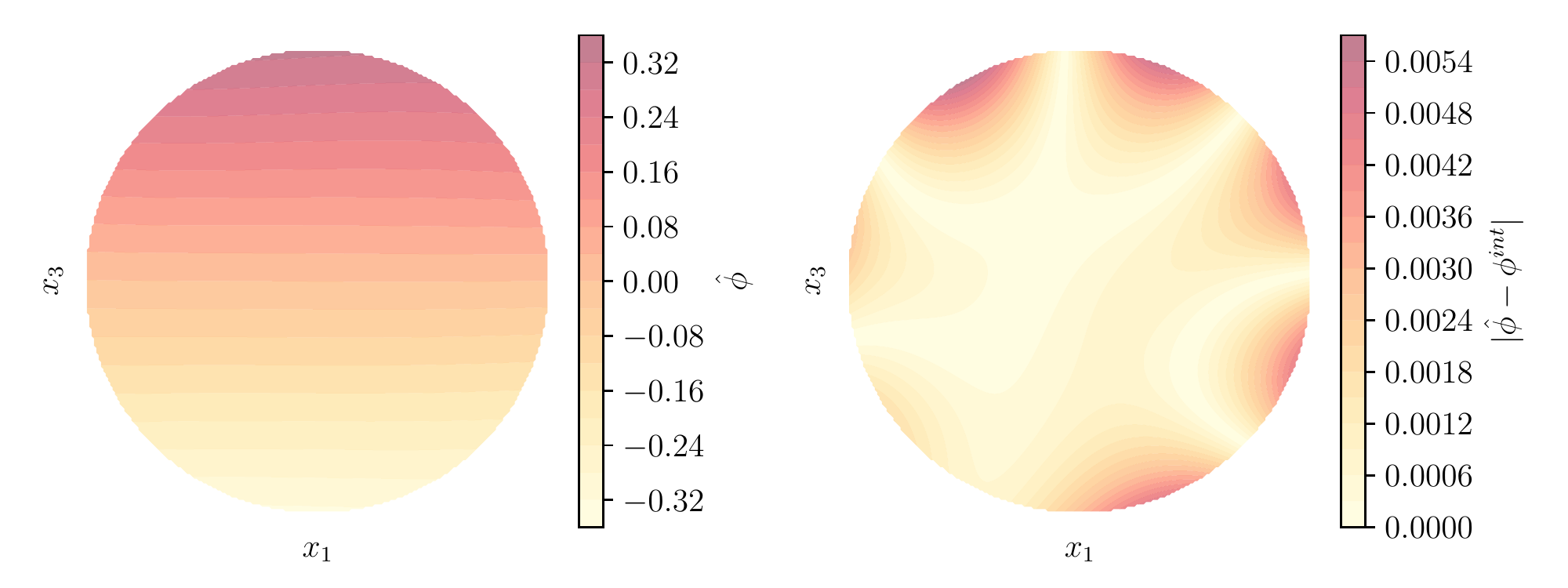}
     \caption{Computed solution to the scalar potential of the stray field of a uniformly magnetized sphere (left), together with the absolute error (right) using ELMs with 16 hidden nodes.}
     \label{fig:stray_field_sphere_16}
 \end{figure}

\begin{figure}[ht]
     \centering
     \includegraphics[width=\textwidth]{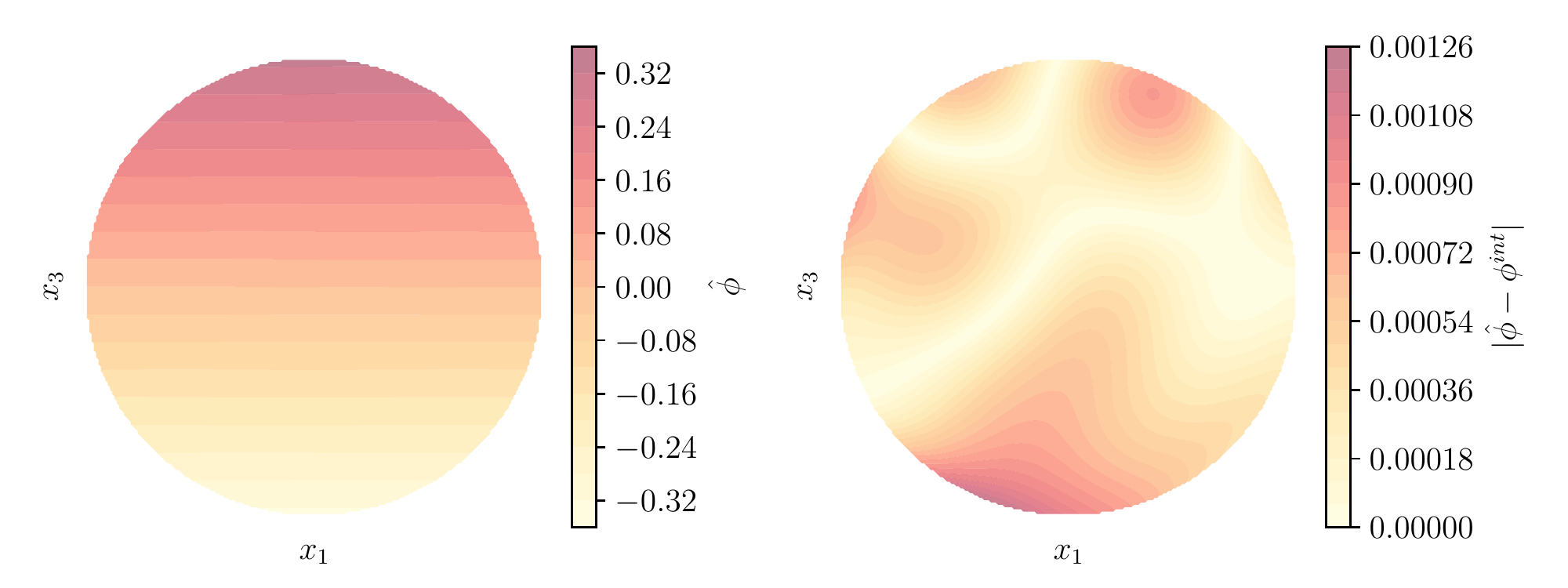}
     \caption{Computed solution to the scalar potential of the stray field of a uniformly magnetized sphere (left), together with the absolute error (right) using ELMs with 32 hidden nodes.}
     \label{fig:stray_field_sphere_32}
 \end{figure}

Figure~\ref{fig:stray_field_sphere_16} shows the solution to the stray field using the ELM method with only 16 hidden nodes per ELM. In total the model has only 48 parameters. We also test the method with 32 hidden nodes which is shown in Figure~\ref{fig:stray_field_sphere_32}. With a higher number of hidden nodes the absolute error decreases. The exact stray field energy is $\pi/9 \approx 0.3491\; [\frac{1}{2}\mu_0M_s^2]$, where our model with 16 hidden nodes has energy $0.3486\; [\frac{1}{2}\mu_0M_s^2]$ and the model with 32 nodes has the energy $0.3487\; [\frac{1}{2}\mu_0M_s^2]$.
 This is of about the same accuracy as the finite element computations in Ref.~\cite{exl2018magnetostatic}, and also shows that the physics-informed approach with hard constraint satisfies the boundary conditions very well. We note that the selection of the initial weights, as well as the choice for the indicator function $\ell$ is important to achieve good accuracy.
 
 The second test case is a unit cube, centered at the origin, with three different magnetization states, a uniform magnetization, a flower state and a vortex state.  
\begin{figure}[!ht]
\centering
\begin{subfigure}{0.4\textwidth}
\centering
\includegraphics[width=\textwidth]{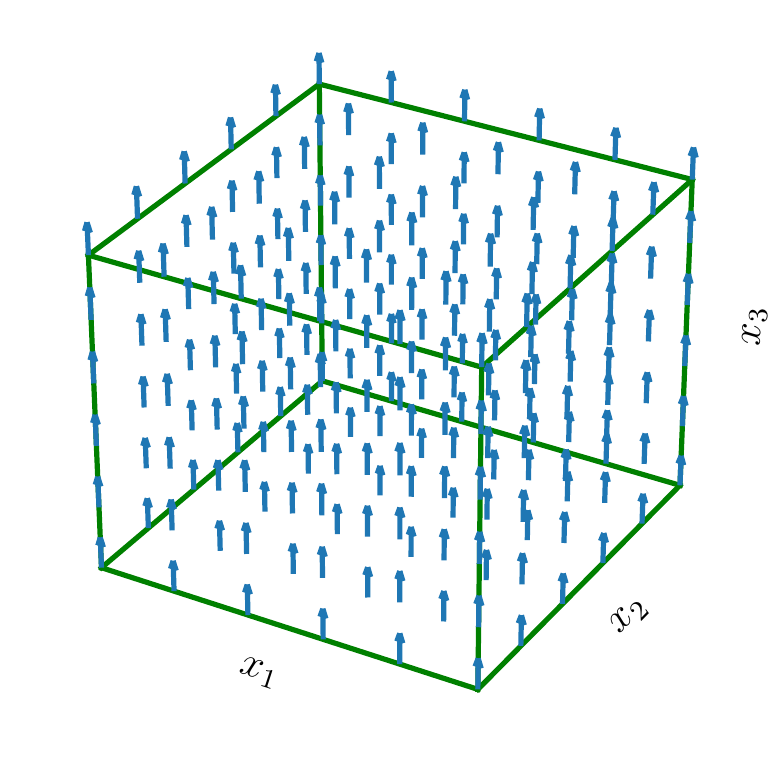}
\end{subfigure}
\begin{subfigure}{0.46\textwidth}
\centering
\raisebox{0.3cm}{\includegraphics[width=\textwidth]{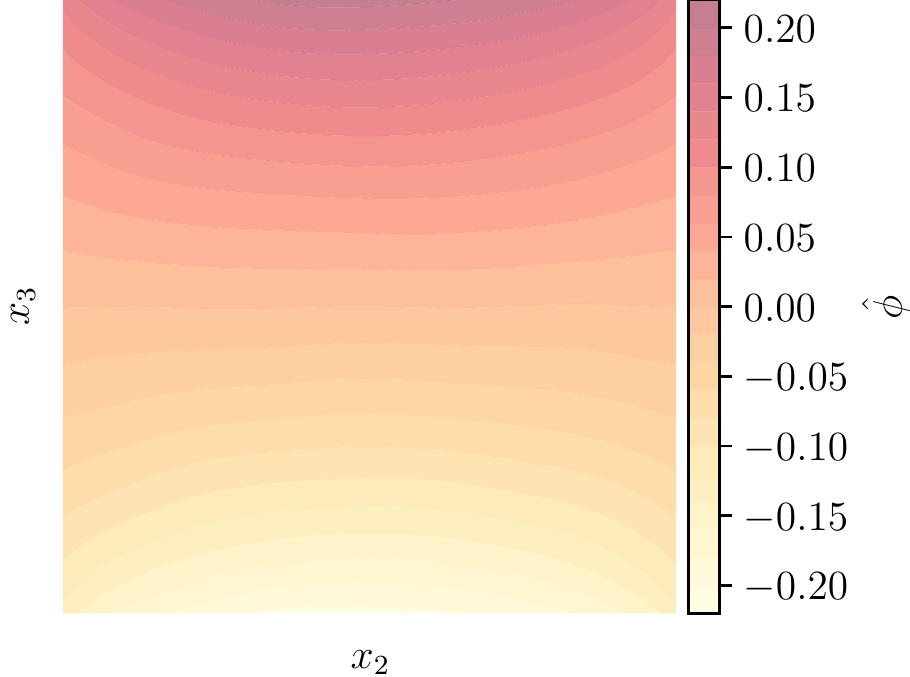}}
\end{subfigure}
\caption{Unit cube with constant magnetization in $x_3$ direction (left) and the stray field potential $\phi$ on the $x_2-x_3$ plane (right). The potential is trained with ELMs with 256 parameters.}
\label{fig:stray_field_const}
\end{figure}

\begin{figure}[!ht]
\centering
\begin{subfigure}{0.4\textwidth}
\centering
\includegraphics[width=\textwidth]{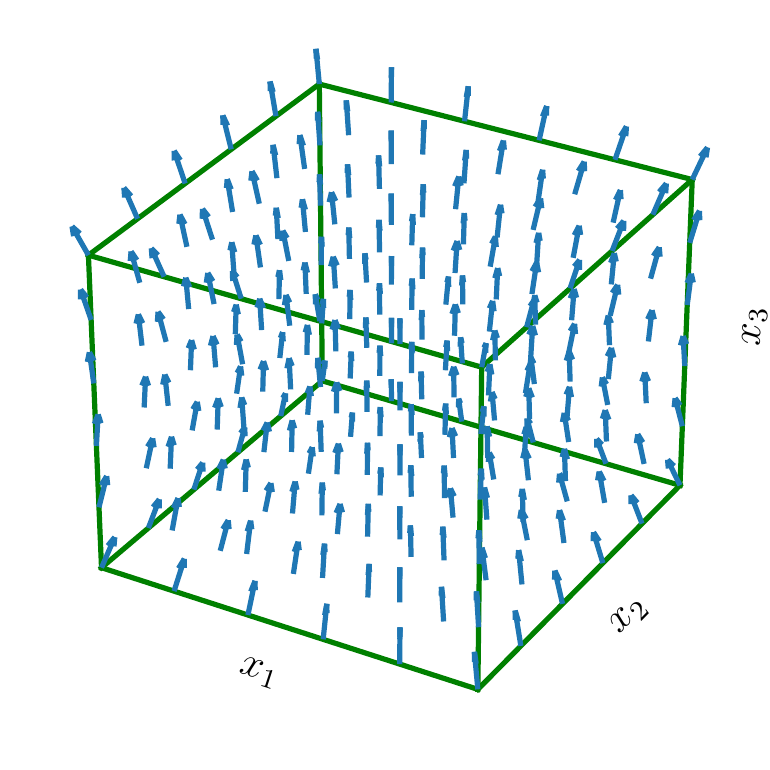}
\end{subfigure}
\begin{subfigure}{0.46\textwidth}
\centering
\raisebox{0.3cm}{\includegraphics[width=\textwidth]{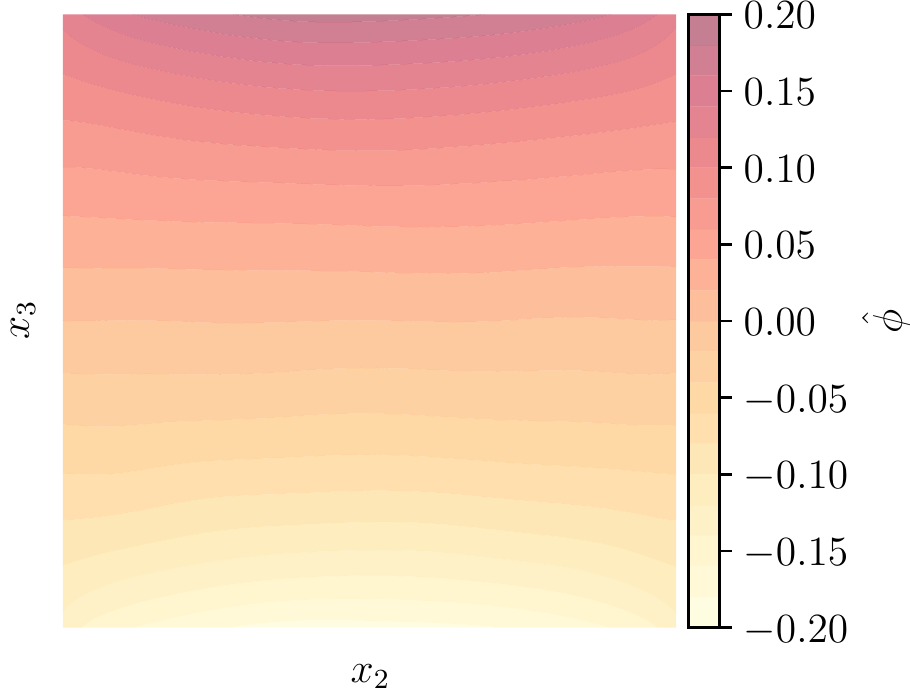}}
\end{subfigure}
\caption{Unit cube with flower state magnetization (left) and the stray field potential $\phi$ on the $x_2-x_3$ plane (right). The potential is trained with ELMs with 256 parameters.}
\label{fig:stray_field_flower}
\end{figure}

\begin{figure}[!ht]
\centering
\begin{subfigure}{0.4\textwidth}
\centering
\includegraphics[width=\textwidth]{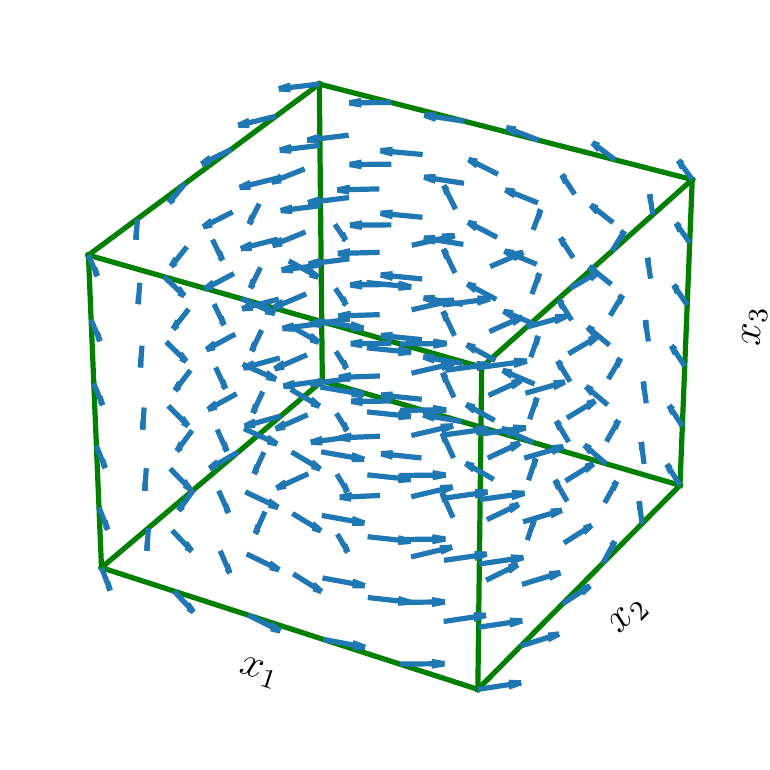}
\end{subfigure}
\begin{subfigure}{0.46\textwidth}
\centering
\raisebox{0.3cm}{\includegraphics[width=\textwidth]{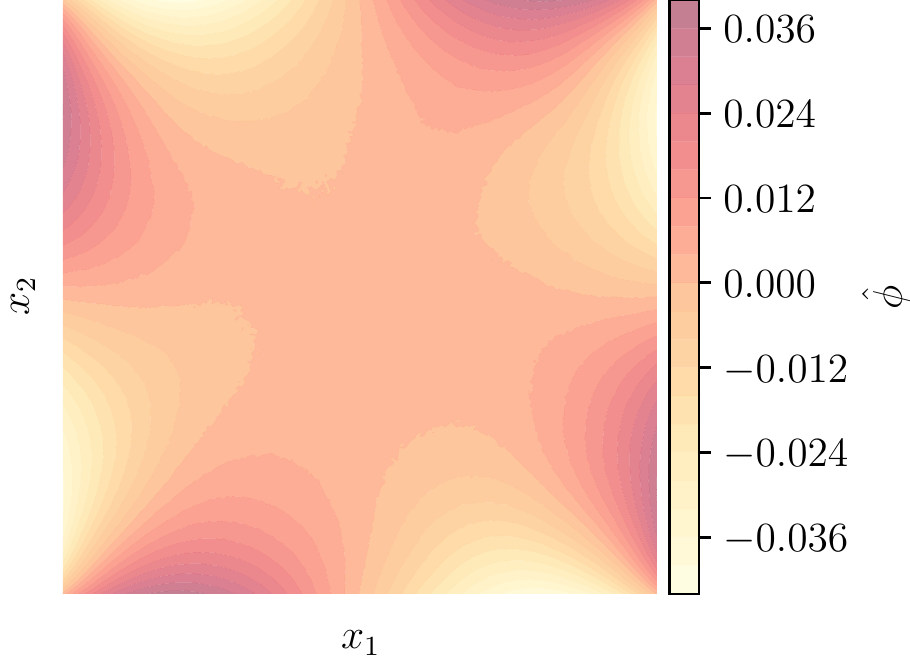}}
\end{subfigure}
\caption{Unit cube with vortex state magnetization (left) and the stray field potential $\phi$ on the $x_1-x_2$ plane (right). The potential is trained with ELMs with 256 parameters.}
\label{fig:stray_field_vortex}
\end{figure}
 The weights are drawn from a $[-4,4]^4$ hypercube and input weights are shared for all ELMs. Each ELM has 256 hidden nodes. We use 4096  sampling points from $\Omega$ and 2048 samples from $\partial\Omega$. Further, for the computation of the boundary integral, 16384 different sample points are used and the integral is approximated with Monte Carlo integration. All sample points are uniformly distributed on the domain and drawn from a Sobol sequence. The indicator function \eqref{eq:indicator_fun} is computed using a sampling method as described in section \ref{sec:stay-field} using 256 sampling points on $\partial\Omega$. Since there is no known exact solution, we can compare the stray field energy to existing solutions in the literature \cite{abert2013numerical, exl2018magnetostatic}, where we take the formulas with the same parameters.

 Figure~\ref{fig:stray_field_const} shows the uniformly magnetized cube on the left and the scalar potential, approximated with ELMs, on the right. The magnetostatic energy is $0.3327\; [\frac{1}{2}\mu_0M_s^2]$, whereas the true value is $1 / 3\; [\frac{1}{2}\mu_0M_s^2]$. Figure~\ref{fig:stray_field_flower} shows the same for the flower state magnetization and we find that the stray field energy is $0.3052\;[\frac{1}{2}\mu_0M_s^2]$. As a reference we use the values computed in Ref.~\cite{abert2013numerical} by the demagnetisation tensor method (DM), which is in this case $0.3160\;[\frac{1}{2}\mu_0M_s^2]$ on a $40 \times 40 \times 40$ grid. Finally, for the vortex state we get an energy of $0.0436\;[\frac{1}{2}\mu_0M_s^2]$ and the reference value is $0.0438\;[\frac{1}{2}\mu_0M_s^2]$. Figure~\ref{fig:stray_field_vortex} shows these results analogue to the uniformly magnetized state and the flower state. All computations are done using \textit{JAX} \cite{jax2018github} on GPU with a \textit{GeForce RTX 2070} using single precision.

\subsection{$\mu$MAG Standard Problem $\#3$}

We will illustrate our approach via the the NIST $\mu$MAG standard problem $\#3$ \cite{mumag3}. The goal is to find the single domain limit of a cubic magnetic particle, that is to find the side length $\lambda$ of the cube of equal energy for the flower and the vortex state. The uniaxial anisotropy is $K_u = 0.1 \,[\tfrac{1}{2} \mu_0 M_s^2]$ with the easy axis directed parallel to the $x_3$-axis of the cube.
The sample set includes the spatial domain sample points as well as the lengths of the cube, i.e. $\mathcal{S}=\{\vec{x}_1,\hdots,\vec{x}_{n_\Omega}\} \times \Lambda$, where $\Lambda= (8,9)$.
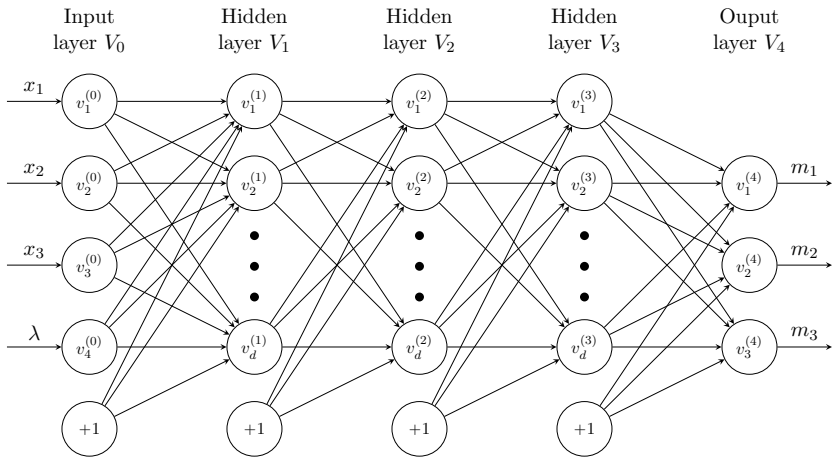
\begin{figure}[!ht]
\centering
\resizebox{0.7\textwidth}{!}{%
	\input{nn_arch.tex}
}
\caption{Neural network architecture for $\vec{m}$ with 3 hidden layers and $d=50$ nodes per hidden layer. There are $5503$ trainable parameters.}
\label{fig:mnet}
\end{figure}
\begin{figure}[ht]
    \centering
    \includegraphics[width=0.8\textwidth]{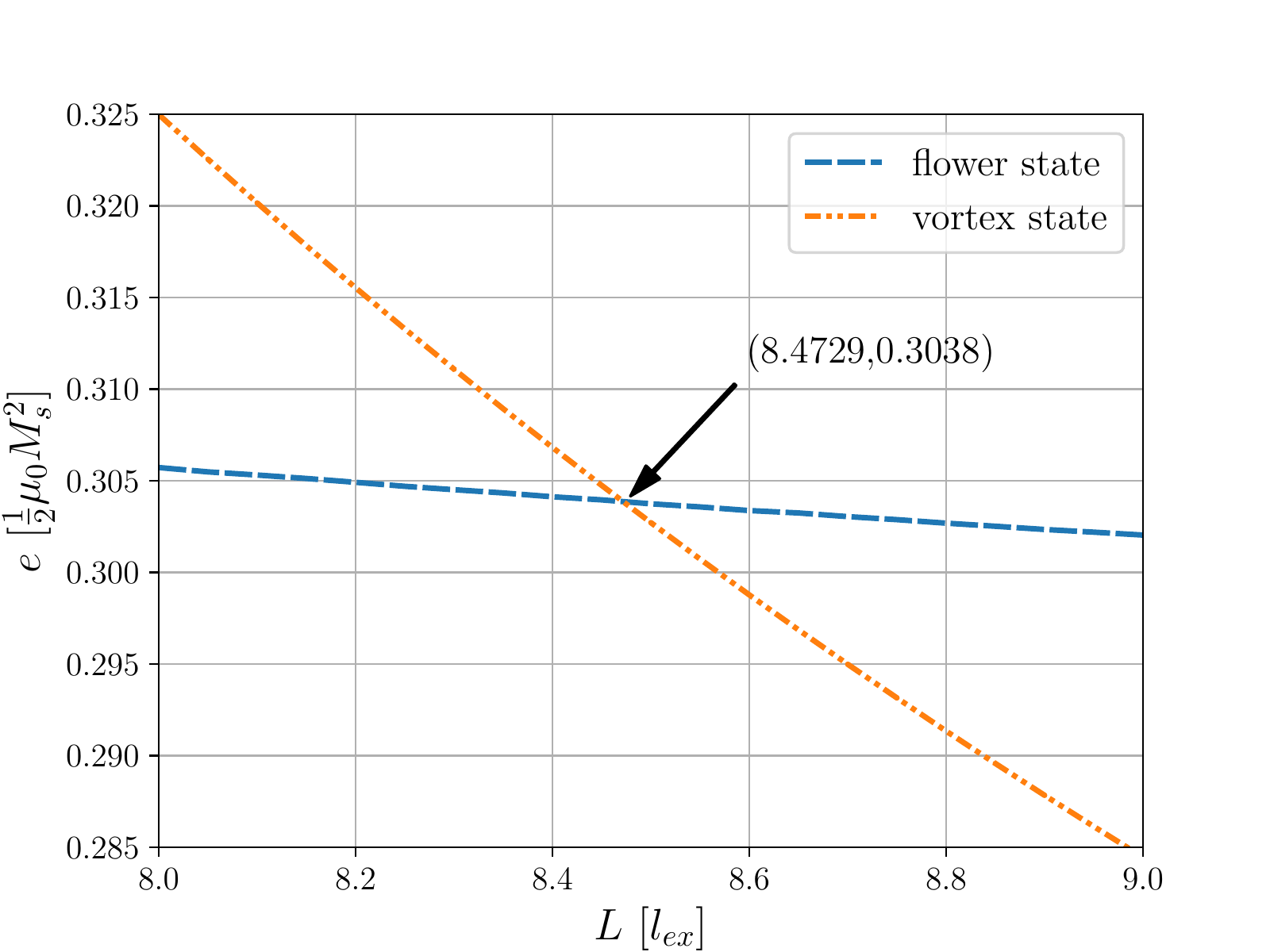}
    \caption{Energy crossing for the $\mu$MAG Standard Problem $\#3$. The Single Domain Limit is at $(8.4729, 0.3038)$. Each line is evaluated from a single low-parametric neural network that represents the local energy optimum at the respective exchange length.}
    \label{fig:crossing}
\end{figure}
The spatial domain sample set $\vec X_{dom}$ includes $4096$ samples which are drawn from a Sobol sequence. We use a batch size of $50$ sample points together with a single instance $\lambda$ of the conditional parameter which is drawn from a uniform distribution. Therefore, for each batch there is a single stray field computation. We use the sample set $\vec X_{dom}$ together with another sample set $\vec X_{bnd}$ with $2048$ samples to train the boundary ELM $g$ in \eqref{eq:hard_cons_formulation}. The solution of the convolution integral \eqref{eq:single-layer-pot} is computed by Monte Carlo sampling with another $16384$ uniform samples. Computation of the stray field requires the training of three ELMs, one for $\phi_1$ and two for $\phi_2$ since the boundary conditions are not homogeneous. Initial weights and biases are also drawn from a Sobol sequence for a hypercube $[-4,4]^4$. We find that those initial weights work well if the unit cube is centered at zero. For fast evaluation, the solution operators are precomputed using \textit{singular value decomposition} (SVD). Each ELM has $256$ hidden nodes and all activation functions are $\tanh$.

We use a \textit{RAdam} optimizer \cite{liu2019variance} with a constant learning rate of $1\mathrm{e}{-5}$ and train the model for $10000$ epochs with the dataset $\vec X_{dom}$. After $1000$ epochs, the penalty parameter $\alpha$ in \eqref{eq:objective_energy_min} is increased by a factor of $1.7$. The initial penalty parameter is $10$. The network architecture for the magnetization is shown in Figure~\ref{fig:mnet}. It is important to set an appropriate initial magnetization for the energy minimization procedure, since there exist multiple local minima. We therefore train the network to approximate an initial magnetization for the flower and vortex state with SGD.

Figure~\ref{fig:crossing} shows the total energy with respect to the cube length $L$ for the optimized flower and the vortex state. Note that each line is evaluated from one single low-parametric neural network and the solutions are continuous. We find that the single domain limit is at $L=8.4729$ with a total reduced energy of $e=0.3038$. Table~\ref{tab:sp3_results} shows the reduced anisotropy energy $e_a$, the reduced exchange energy $e_{ex}$, the reduced magnetostatic energy $e_s$ as well as the mean magnetization for the single domain limit. 
\begin{table}[!ht]
    \centering
    \begin{tabular}{l|c c|c c}
        & Flower State & & Vortex State &\\
        \hline
        $e_a$ & 0.0052 & (0.0052) & 0.0519 & (0.0522)\\
        $e_{ex}$ & 0.0159 & (0.0158) & 0.1724 & (0.1696)\\
        $e_s$ & 0.2827 & (0.2839) & 0.0795 & (0.0830)\\
        $\left<m_1\right>$ & 0.0078 & (-) & 0.0107 & (-)\\
        $\left<m_2\right>$ & 0.0005 & (-) & 0.3461 & (0.351)\\
        $\left<m_3\right>$ & 0.9733 & (0.973) & 0.0086 & (-)\\
        $\|\err_c\|_{\max}$ & 0.0040 & (-) & 0.0062 & (-)\\
        $\mathrm{MSE}(\err_c)$ & $9.2\mathrm{e}{-07}$ & (-) & $1.4\mathrm{e}{-06}$ & (-)\\
    \end{tabular}
    \caption{Energies, mean magnetizations and constraint violation in the single domain limit. The values in brackets are here stated for reference and were previously published in \cite{hertel2002finite}.}
    \label{tab:sp3_results}
\end{table}
Here $\err_c(\vec x) = \|\vec m(\vec x)\| - 1$ is the norm constraint violation. The values are in excellent agreement with previously published results \cite{mumag3}.

\begin{figure}[!ht]
    \centering
    \includegraphics[width=0.8\textwidth]{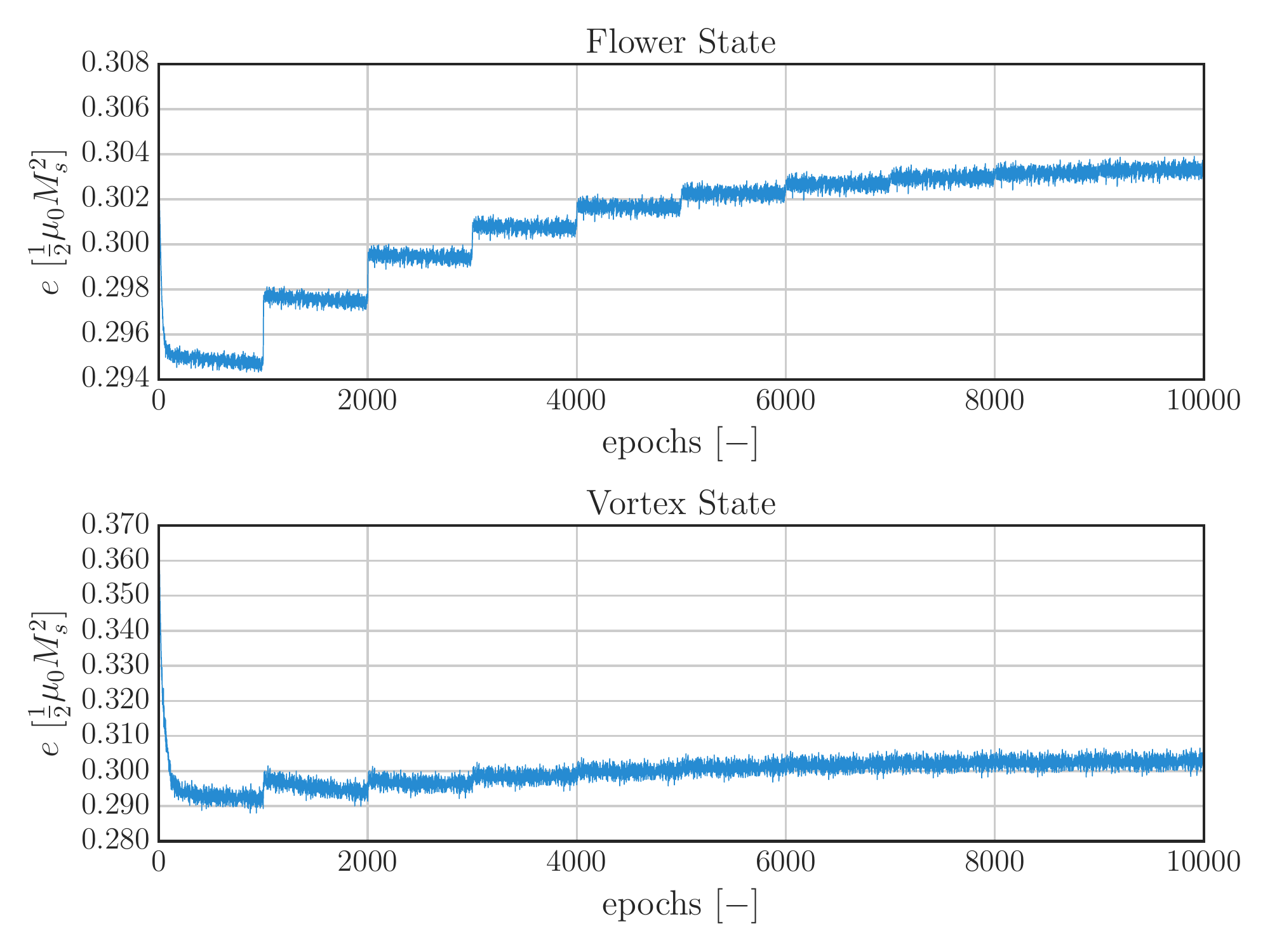}
    \caption{Learning curve for the total Gibbs free energy for the flower and the vortex state.}
    \label{fig:lc_energy}
\end{figure}

\begin{figure}[!ht]
    \centering
    \includegraphics[width=0.8\textwidth]{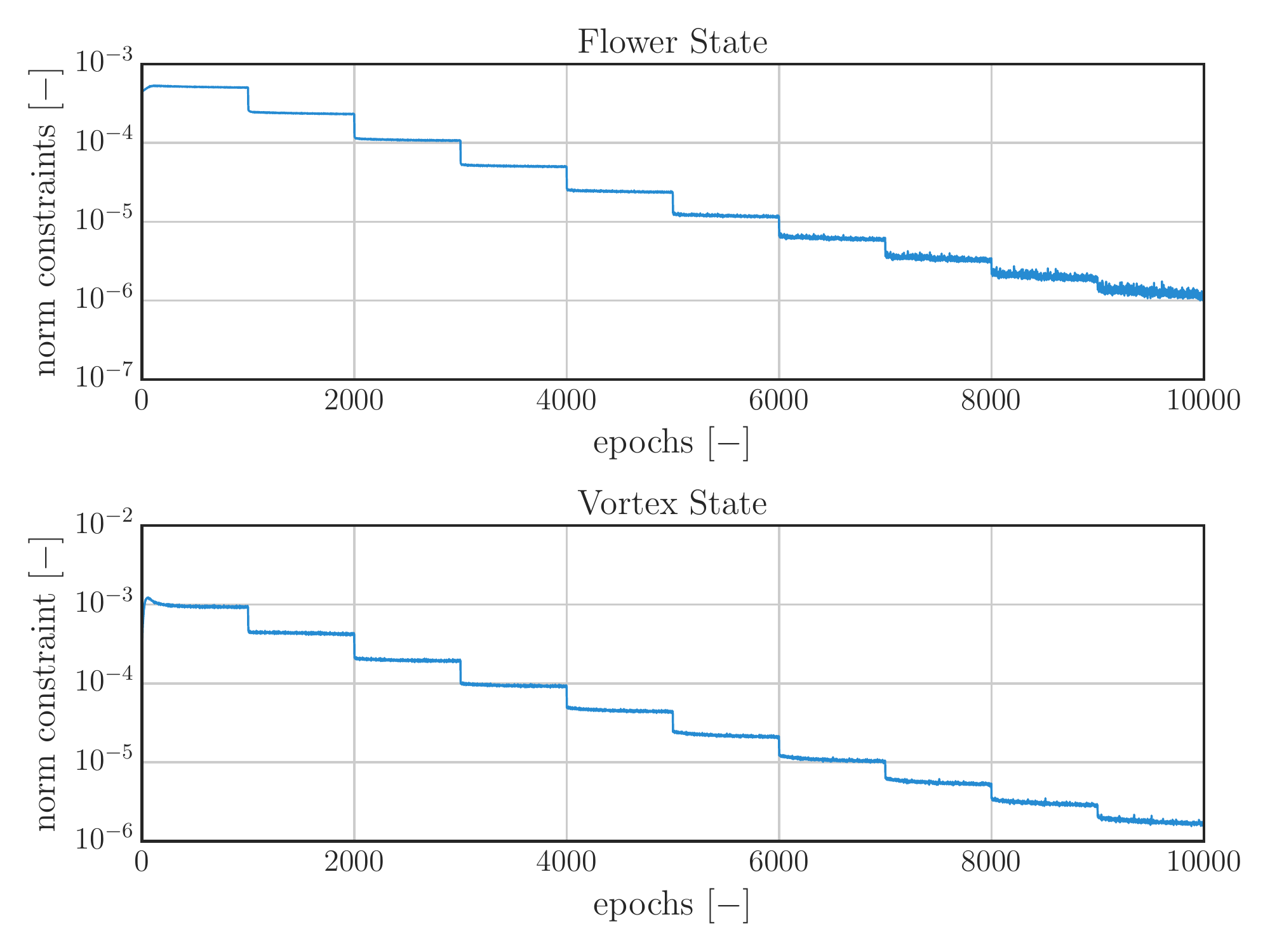}
    \caption{Norm constraint violation during training for flower and vortex state.}
    \label{fig:norm_con_violation}
\end{figure}

The training curves for the total energy (Figure~\ref{fig:lc_energy}) and the norm constraint violation (Figure~\ref{fig:norm_con_violation}) show the effect of the increasing penalty term in \eqref{eq:objective_energy_min}. As the penalty parameter scaled up, the norm constraint violation is punished more severely and the energy increases. It is important that the initial penalty is not too small, since this would make it easier during training to jump out of a local minimum and converge to another state. We note here that the total energy in Figure~\ref{fig:lc_energy} is the mean over the domain of $\Lambda$.


\section{Conclusion}
We performed full 3d static micromagnetic computations using a physics-informed neural networks for the magnetization configuration, where we especially presented a novel approach for the stray field computation. This approach is unsupervised, mesh-free and allows to consider additional model parameters such as the exchange length. The involved stray field problem is solved with a simple linear shallow neural network ansatz via extreme learning machines (ELM) in a splitting approach for the magnetic scalar potential. ELMs are surprisingly simple, in essence only consisting of a linear combination of nonlinear activation functions fulfilling an universal approximation property similar to deep neural networks. We use a penalty-free physics-informed ELM approach, which comes with several advantages, e.g. an exact known solution operator obtained from a linear least squares formulation and exact fulfillment of the boundary conditions. We show how the stray field problem can be solved in a sequential way, where basically only the solutions of Dirichlet Poisson problems need to be learned.

We justify our methods by means of numerical examples both for the stray field computation as well as for the full 3d micromagnetic energy minimization. The latter is tested for the $\mu$MAG standard problem $\#3$ where the single domain limit of a small cubic particle is derived from two single neural networks modelling the whole parameter range of the exchange length. Our findings are in excellent agreement with previously published results obtained from conventional mesh-based methods, but represent continuous interpolations of the solutions in the entire considered parameter range.

Future work will aim to improve the existing approach by considering more elaborate optimization methods in the training process, importance sampling for the stray field and domain decomposition to generalize to more complex magnetic domains. This will allow the application of the method to e.g. permanent magnet grain models as well as simple data-fusion with experimental data owing to the mesh-free character of our proposed methods. Furthermore, we are currently working on a shortcut of the presented stray field learning algorithm via a previously published novel stray field energy formula which has the ability to reduce the necessary amount of ELM models to just a single one \cite{exl2018magnetostatic}. Future work will also consider unsupervised learning of the solution to the Landau-Lifschitz-Gilbert equation improving our recent works on supervised learning for magnetization dynamics \cite{kovacs2019learning,schaffer2021machine}.

\section*{Acknowledgements}

\noindent Financial support by the Austrian Science Fund (FWF) via project P 31140 ”Reduced Order Approaches for Micromagnetics (ROAM)” and project P 35413 ”Design of Nanocomposite Magnets by
Machine Learning (DeNaMML)” is gratefully acknowledged. The authors acknowledge the University of Vienna research platform MMM Mathematics - Magnetism - Materials. The computations were partly achieved by using the Vienna Scientific Cluster (VSC) via the funded projects No. 71140 and 71952.
This research was funded in whole or in part by the Austrian Science Fund (FWF) [P 31140, P 35413]. For the purpose of Open Access, the author has applied a CC BY public copyright licence to any Author Accepted Manuscript (AAM) version arising from this submission.

%%%%%%%%%%%%%%%%%%%%
\bibliographystyle{abbrv}
\bibliography{bibref}

\end{document}

%% file: elm.tex
\tikzset{%
	every neuron/.style={
		circle,
		draw,
		minimum size=1cm
	},
	neuron missing/.style={
		draw=none, 
		scale=4,
		text height=0.333cm,
		execute at begin node=\color{black}$\vdots$
	},
	gap/.style={
		draw=none, 
		fill=none
	},
}

\begin{tikzpicture}[x=1.5cm, y=1.5cm, >=stealth]

\foreach \m [count=\y] in {1,2,3,4}
	\node [every neuron/.try, neuron \m/.try ] (input-\m) at (0,4.5-\y) {};

\foreach \m [count=\y] in {1,2,missing,3,4}
	\node [every neuron/.try, neuron \m/.try ] (hidden-\m) at (2,5-\y) {};

%\foreach \m [count=\y] in {1,2,missing,3,4}
%	\node [every neuron/.try, neuron \m/.try ] (hidden2-\m) at (4,5-\y) {};

%\node [draw=none,fill=none,scale=3] () at (3.1,2) {$\cdots$};

% \foreach \m [count=\y] in {1,2,missing,3,4}
% 	\node [every neuron/.try, neuron \m/.try ] (hidden3-\m) at (6,5-\y) {};
	
\foreach \m [count=\y] in {1}
	\node [every neuron/.try, neuron \m/.try ] (output-\m) at (4,3-\y) {};

\foreach \l [count=\i] in {1,2,3}
	\draw [<-] (input-\i) -- ++(-1,0)
	node [above, midway] {$x_\l$};

\foreach \l [count=\i] in {$\,$,$\,$,$\,$,$+1$}
	\node [scale=0.8] at (input-\i) {\l};

\foreach \l [count=\i] in {$\sigma_1$,$\sigma_2$,$\sigma_{N-1}$,$\sigma_{N}$}
	\node [scale=0.8] at (hidden-\i) {\l};

%\foreach \l [count=\i] in {$v^{(T-2)}_1$,$v^{(T-2)}_2$,$v^{(T-2)}_{N_{T-2}}$,$+1$}
%	\node [scale=0.8] at (hidden2-\i) {\l};

% \foreach \l [count=\i] in {$v^{(T-1)}_1$,$v^{(T-1)}_2$,$v^{(T-1)}_{N_{T-1}}$,$+1$}
% 	\node [scale=0.8] at (hidden3-\i) {\l};

\foreach \l [count=\i] in {$+$}
	\node [scale=0.8] at (output-\i) {\l};

\foreach \l [count=\i] in {1}
	\draw [->] (output-\i) -- ++(1,0)
	node [above, midway] {$\hat \phi$};

\foreach \i in {1,...,4}
\foreach \j in {1,...,4}
	\draw [->] (input-\i) -- (hidden-\j);

% \foreach \i in {1,...,4}
% \foreach \j in {1,...,3}
% 	\draw [->] (hidden-\i) -- (hidden3-\j);

\foreach \l [count=\i]  in {$\;\;\beta_1$,$\beta_2$,$\beta_{N-1}\;\;\;$,$\beta_N\;\;$}%{1,...,4}
\foreach \j in {1}
\draw [->] (hidden-\i) -- node[above] {\l} (output-\j);

\foreach \l [count=\x from 0] in {Input \\ layer, Hidden \\ layer, Ouput \\ layer}
\node [align=center, above] at (\x*2,4.5) {\l};

\end{tikzpicture}

%% file: nn_arch.tex
\tikzset{%
	every neuron/.style={
		circle,
		draw,
		minimum size=1cm
	},
	neuron missing/.style={
		draw=none, 
		scale=4,
		text height=0.333cm,
		execute at begin node=\color{black}$\vdots$
	},
	gap/.style={
		draw=none, 
		fill=none
	},
}

\begin{tikzpicture}[x=1.5cm, y=1.5cm, >=stealth]

\foreach \m [count=\y] in {1,2,3,4,5}
	\node [every neuron/.try, neuron \m/.try ] (input-\m) at (0,5-\y) {};

\foreach \m [count=\y] in {1,2,missing,3,4}
	\node [every neuron/.try, neuron \m/.try ] (hidden-\m) at (2,5-\y) {};

\foreach \m [count=\y] in {1,2,missing,3,4}
	\node [every neuron/.try, neuron \m/.try ] (hidden2-\m) at (4,5-\y) {};

\foreach \m [count=\y] in {1,2,missing,3,4}
	\node [every neuron/.try, neuron \m/.try ] (hidden3-\m) at (6,5-\y) {};
	
\foreach \m [count=\y] in {1,2,3}
	\node [every neuron/.try, neuron \m/.try ] (output-\m) at (8,4-\y) {};

\foreach \l [count=\i] in {$x_1$,$x_2$,$x_3$,$\lambda$}
	\draw [<-] (input-\i) -- ++(-1,0)
	node [above, midway] {\l};

\foreach \l [count=\i] in {$v^{(0)}_1$,$v^{(0)}_2$,$v^{(0)}_3$,$v^{(0)}_4$,$+1$}
	\node [scale=0.8] at (input-\i) {\l};

\foreach \l [count=\i] in {$v^{(1)}_1$,$v^{(1)}_2$,$v^{(1)}_{d}$,$+1$}
	\node [scale=0.8] at (hidden-\i) {\l};

\foreach \l [count=\i] in {$v^{(2)}_1$,$v^{(2)}_2$,$v^{(2)}_{d}$,$+1$}
	\node [scale=0.8] at (hidden2-\i) {\l};

\foreach \l [count=\i] in {$v^{(3)}_1$,$v^{(3)}_2$,$v^{(3)}_{d}$,$+1$}
	\node [scale=0.8] at (hidden3-\i) {\l};

\foreach \l [count=\i] in {$v^{(4)}_1$,$v^{(4)}_2$,$v^{(4)}_3$}
	\node [scale=0.8] at (output-\i) {\l};

\foreach \l [count=\i] in {1,2,3}
	\draw [->] (output-\i) -- ++(1,0)
	node [above, midway] {$m_{\l}$};

\foreach \i in {1,...,5}
\foreach \j in {1,...,3}
	\draw [->] (input-\i) -- (hidden-\j);

\foreach \i in {1,...,4}
\foreach \j in {1,...,3}
	\draw [->] (hidden-\i) -- (hidden2-\j);

\foreach \i in {1,...,4}
\foreach \j in {1,...,3}
	\draw [->] (hidden2-\i) -- (hidden3-\j);

\foreach \i in {1,...,4}
\foreach \j in {1,...,3}
\draw [->] (hidden3-\i) -- (output-\j);

\foreach \l [count=\x from 0] in {Input \\ layer $V_0$, Hidden \\ layer $V_1$, Hidden \\ layer $V_2$, Hidden \\ layer $V_3$, Ouput \\ layer $V_4$}
\node [align=center, above] at (\x*2,4.5) {\l};

\end{tikzpicture}

%% file: main.bbl
\begin{thebibliography}{10}

\bibitem{abert2013numerical}
C.~Abert, L.~Exl, G.~Selke, A.~Drews, and T.~Schrefl.
\newblock Numerical methods for the stray-field calculation: {A} comparison of
  recently developed algorithms.
\newblock {\em Journal of Magnetism and Magnetic Materials}, 326:176--185,
  2013.

\bibitem{bashir2012head}
M.~Bashir, T.~Schrefl, J.~Dean, A.~Goncharov, G.~Hrkac, D.~Allwood, and
  D.~Suess.
\newblock Head and bit patterned media optimization at areal densities of 2.5
  {T}bit/in2 and beyond.
\newblock {\em Journal of Magnetism and Magnetic Materials}, 324(3):269--275,
  2012.

\bibitem{bjork2021magtense}
R.~Bj{\o}rk, E.~B. Poulsen, K.~K. Nielsen, and A.~R. Insinga.
\newblock Magtense: A micromagnetic framework using the analytical
  demagnetization tensor.
\newblock {\em Journal of Magnetism and Magnetic Materials}, 535:168057, 2021.

\bibitem{jax2018github}
J.~Bradbury, R.~Frostig, P.~Hawkins, M.~J. Johnson, C.~Leary, D.~Maclaurin,
  G.~Necula, A.~Paszke, J.~Vander{P}las, S.~Wanderman-{M}ilne, and Q.~Zhang.
\newblock {JAX}: composable transformations of {P}ython+{N}um{P}y programs,
  2018.
\newblock \url{http://github.com/google/jax}.

\bibitem{carstensen1995adaptive}
C.~Carstensen and E.~P. Stephan.
\newblock Adaptive coupling of boundary elements and finite elements.
\newblock {\em ESAIM: Mathematical Modelling and Numerical Analysis},
  29(7):779--817, 1995.

\bibitem{d2009spectral}
M.~d'Aquino, C.~Serpico, G.~Bertotti, T.~Schrefl, and I.~Mayergoyz.
\newblock Spectral micromagnetic analysis of switching processes.
\newblock {\em Journal of Applied Physics}, 105(7):07D540, 2009.

\bibitem{exl2014tensor}
L.~Exl.
\newblock {\em Tensor grid methods for micromagnetic simulations}.
\newblock PhD thesis, 2014.

\bibitem{exl2018magnetostatic}
L.~Exl.
\newblock A magnetostatic energy formula arising from the {L}2-orthogonal
  decomposition of the stray field.
\newblock {\em Journal of Mathematical Analysis and Applications},
  467(1):230--237, 2018.

\bibitem{exl2014labonte}
L.~Exl, S.~Bance, F.~Reichel, T.~Schrefl, H.~Peter~Stimming, and N.~J. Mauser.
\newblock La{B}onte's method revisited: An effective steepest descent method
  for micromagnetic energy minimization.
\newblock {\em Journal of Applied Physics}, 115(17):17D118, 2014.

\bibitem{exl2019preconditioned}
L.~Exl, J.~Fischbacher, A.~Kovacs, H.~Oezelt, M.~Gusenbauer, and T.~Schrefl.
\newblock Preconditioned nonlinear conjugate gradient method for micromagnetic
  energy minimization.
\newblock {\em Computer Physics Communications}, 235:179--186, 2019.

\bibitem{exl2014non}
L.~Exl and T.~Schrefl.
\newblock Non-uniform {FFT} for the finite element computation of the
  micromagnetic scalar potential.
\newblock {\em Journal of Computational Physics}, 270:490--505, 2014.

\bibitem{exl2020micromagnetism}
L.~Exl, D.~Suess, and T.~Schrefl.
\newblock Micromagnetism.
\newblock {\em Handbook of Magnetism and Magnetic Materials}, pages 1--44,
  2020.

\bibitem{fischbacher2018micromagnetics}
J.~Fischbacher, A.~Kovacs, M.~Gusenbauer, H.~Oezelt, L.~Exl, S.~Bance, and
  T.~Schrefl.
\newblock Micromagnetics of rare-earth efficient permanent magnets.
\newblock {\em Journal of Physics D: Applied Physics}, 51(19):193002, 2018.

\bibitem{fischbacher2017nonlinear}
J.~Fischbacher, A.~Kovacs, H.~Oezelt, T.~Schrefl, L.~Exl, J.~Fidler, D.~Suess,
  N.~Sakuma, M.~Yano, A.~Kato, et~al.
\newblock Nonlinear conjugate gradient methods in micromagnetics.
\newblock {\em AIP Advances}, 7(4):045310, 2017.

\bibitem{garcia2006adaptive}
C.~J. Garcia-Cervera and A.~M. Roma.
\newblock Adaptive mesh refinement for micromagnetics simulations.
\newblock {\em Magnetics, IEEE Transactions on}, 42(6):1648--1654, 2006.

\bibitem{hertel2002finite}
R.~Hertel and H.~Kronm{\"u}ller.
\newblock Finite element calculations on the single-domain limit of a
  ferromagnetic cube—a solution to $\mu${MAG} {S}tandard {P}roblem {N}o. 3.
\newblock {\em Journal of magnetism and magnetic materials}, 238(2-3):185--199,
  2002.

\bibitem{hornik1989multilayer}
K.~Hornik, M.~Stinchcombe, and H.~White.
\newblock Multilayer feedforward networks are universal approximators.
\newblock {\em Neural networks}, 2(5):359--366, 1989.

\bibitem{huang2006universal}
G.-B. Huang, L.~Chen, C.~K. Siew, et~al.
\newblock Universal approximation using incremental constructive feedforward
  networks with random hidden nodes.
\newblock {\em IEEE Trans. Neural Networks}, 17(4):879--892, 2006.

\bibitem{huang2006extreme}
G.-B. Huang, Q.-Y. Zhu, and C.-K. Siew.
\newblock Extreme learning machine: theory and applications.
\newblock {\em Neurocomputing}, 70(1-3):489--501, 2006.

\bibitem{huber20173d}
C.~Huber, C.~Abert, F.~Bruckner, M.~Groenefeld, S.~Schuschnigg, I.~Teliban,
  C.~Vogler, G.~Wautischer, R.~Windl, and D.~Suess.
\newblock 3d printing of polymer-bonded rare-earth magnets with a variable
  magnetic compound fraction for a predefined stray field.
\newblock {\em Scientific reports}, 7(1):1--8, 2017.

\bibitem{mumag3}
A.~Hubert and R.~McMichael.
\newblock $\mu${MAG} {Standard} {Problem} \#3.
\newblock \url{https://www.ctcms.nist.gov/~rdm/mumag.org.html}.

\bibitem{kovacs2022magnetostatics}
A.~Kovacs, L.~Exl, A.~Kornell, J.~Fischbacher, M.~Hovorka, M.~Gusenbauer,
  L.~Breth, H.~Oezelt, D.~Praetorius, D.~Suess, et~al.
\newblock Magnetostatics and micromagnetics with physics informed neural
  networks.
\newblock {\em Journal of Magnetism and Magnetic Materials}, 548:168951, 2022.

\bibitem{kovacs2022conditional}
A.~Kovacs, L.~Exl, A.~Kornell, J.~Fischbacher, M.~Hovorka, M.~Gusenbauer,
  L.~Breth, H.~Oezelt, M.~Yano, N.~Sakuma, et~al.
\newblock Conditional physics informed neural networks.
\newblock {\em Communications in Nonlinear Science and Numerical Simulation},
  104:106041, 2022.

\bibitem{kovacs2019learning}
A.~Kovacs, J.~Fischbacher, H.~Oezelt, M.~Gusenbauer, L.~Exl, F.~Bruckner,
  D.~Suess, and T.~Schrefl.
\newblock Learning magnetization dynamics.
\newblock {\em Journal of Magnetism and Magnetic Materials}, 491:165548, 2019.

\bibitem{liu2019variance}
L.~Liu, H.~Jiang, P.~He, W.~Chen, X.~Liu, J.~Gao, and J.~Han.
\newblock On the variance of the adaptive learning rate and beyond.
\newblock {\em arXiv preprint arXiv:1908.03265}, 2019.

\bibitem{miltat2007numerical}
J.~E. Miltat and M.~J. Donahue.
\newblock Numerical micromagnetics: Finite difference methods.
\newblock {\em Handbook of magnetism and advanced magnetic materials}, 2007.

\bibitem{mortari2017theory}
D.~Mortari.
\newblock The theory of connections: {C}onnecting points.
\newblock {\em Mathematics}, 5(4):57, 2017.

\bibitem{nocedal1999numerical}
J.~Nocedal and S.~J. Wright.
\newblock {\em Numerical optimization}.
\newblock Springer, 1999.

\bibitem{perna2022computational}
S.~Perna, F.~Bruckner, C.~Serpico, D.~Suess, and M.~d’Aquino.
\newblock Computational micromagnetics based on normal modes: Bridging the gap
  between macrospin and full spatial discretization.
\newblock {\em Journal of Magnetism and Magnetic Materials}, 546:168683, 2022.

\bibitem{raissi2019physics}
M.~Raissi, P.~Perdikaris, and G.~E. Karniadakis.
\newblock Physics-informed neural networks: A deep learning framework for
  solving forward and inverse problems involving nonlinear partial differential
  equations.
\newblock {\em Journal of Computational physics}, 378:686--707, 2019.

\bibitem{schaffer2021machine}
S.~Schaffer, N.~J. Mauser, T.~Schrefl, D.~Suess, and L.~Exl.
\newblock Machine learning methods for the prediction of micromagnetic
  magnetization dynamics.
\newblock {\em IEEE Transactions on Magnetics}, 58(2):1--6, 2021.

\bibitem{schrefl2007numerical}
T.~Schrefl, G.~Hrkac, S.~Bance, D.~Suess, O.~Ertl, and J.~Fidler.
\newblock Numerical methods in micromagnetics (finite element method).
\newblock {\em Handbook of magnetism and advanced magnetic materials}, 2007.

\bibitem{sheng2021pfnn}
H.~Sheng and C.~Yang.
\newblock Pfnn: A penalty-free neural network method for solving a class of
  second-order boundary-value problems on complex geometries.
\newblock {\em Journal of Computational Physics}, 428:110085, 2021.

\bibitem{smith2010demagnetizing}
A.~Smith, K.~K. Nielsen, D.~Christensen, C.~R.~H. Bahl, R.~Bj{\o}rk, and
  J.~Hattel.
\newblock The demagnetizing field of a nonuniform rectangular prism.
\newblock {\em Journal of Applied Physics}, 107(10):103910, 2010.

\bibitem{sobol1976uniformly}
I.~M. Sobol.
\newblock Uniformly distributed sequences with an additional uniform property.
\newblock {\em USSR Computational Mathematics and Mathematical Physics},
  16(5):236--242, 1976.

\bibitem{suess2018topologically}
D.~Suess, A.~Bachleitner-Hofmann, A.~Satz, H.~Weitensfelder, C.~Vogler,
  F.~Bruckner, C.~Abert, K.~Pr{\"u}gl, J.~Zimmer, C.~Huber, et~al.
\newblock Topologically protected vortex structures for low-noise magnetic
  sensors with high linear range.
\newblock {\em Nature Electronics}, 1(6):362--370, 2018.

\bibitem{wright2003radial}
G.~B. Wright.
\newblock {\em Radial basis function interpolation: numerical and analytical
  developments}.
\newblock University of Colorado at Boulder, 2003.

\end{thebibliography}
